\documentclass{article}

\PassOptionsToPackage{colorlinks=true, citecolor=blue, linkcolor=blue, urlcolor=black}{hyperref}

\usepackage{arxiv}
\usepackage[utf8]{inputenc}
\usepackage[T1]{fontenc}
\usepackage{url}
\usepackage{booktabs}
\usepackage{nicefrac}
\usepackage{microtype}
\usepackage{lipsum}
\usepackage{graphicx}
\usepackage{natbib}
\usepackage{doi}
\usepackage{algorithm}
\usepackage{algorithmicx}
\usepackage{multirow}
\usepackage{multicol}
\usepackage{amsmath,amssymb,amsfonts}
\usepackage{amsthm}
\usepackage{mathrsfs}
\usepackage{xcolor}
\usepackage{float}

\title{Seismic wavefield solutions via physics-guided generative neural operator}

\author{ \href{https://orcid.org/0000-0001-8868-7967}{\includegraphics[scale=0.06]{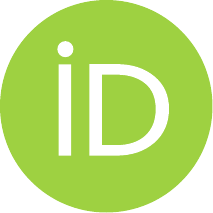}\hspace{1mm}Shijun~Cheng}\\
	Division of Physical Science and Engineering\\
	King Abdullah University of Science and Technology\\
	Thuwal 23955-6900, Saudi Arabia \\
	\texttt{sjcheng.academic@gmail.com} \\
        \And
	\href{https://orcid.org/0000-0002-6336-7614}{\includegraphics[scale=0.06]{orcid.pdf}\hspace{1mm}Mohammad H. Taufik} \\
	Division of Physical Science and Engineering\\
	King Abdullah University of Science and Technology\\
	Thuwal 23955-6900, Saudi Arabia \\
	\texttt{mohammad.taufik@kaust.edu.sa} \\
        \And
	\href{https://orcid.org/0000-0002-9363-9799}{\includegraphics[scale=0.06]{orcid.pdf}\hspace{1mm}Tariq~Alkhalifah} \\
	Division of Physical Science and Engineering\\
	King Abdullah University of Science and Technology\\
	Thuwal 23955-6900, Saudi Arabia \\
	\texttt{tariq.alkhalifah@kaust.edu.sa} \\
}

\renewcommand{\shorttitle}{Physics-guided generative neural operator}

\hypersetup{
pdftitle={A template for the arxiv style},
pdfsubject={q-bio.NC, q-bio.QM},
pdfauthor={David S.~Hippocampus, Elias D.~Striatum},
pdfkeywords={First keyword, Second keyword, More},
}

\begin{document}
\maketitle

\begin{abstract}
Current neural operators often struggle to generalize to complex, out-of-distribution conditions, limiting their ability in seismic wavefield representation. To address this, we propose a generative neural operator (GNO) that leverages generative diffusion models (GDMs) to learn the underlying statistical distribution of scattered wavefields while incorporating a physics-guided sampling process at each inference step. This physics guidance enforces wave equation-based constraints corresponding to specific velocity models, driving the iteratively generated wavefields toward physically consistent solutions. By training the diffusion model on wavefields corresponding to a diverse dataset of velocity models, frequencies, and source positions, our GNO enables to rapidly synthesize high-fidelity wavefields at inference time. Numerical experiments demonstrate that our GNO not only produces accurate wavefields matching numerical reference solutions, but also generalizes effectively to previously unseen velocity models and frequencies. 
\end{abstract}


\section{\textbf{Introduction}}
Traditionally, forward wavefield simulations rely on numerical solution schemes, such as finite-difference \citep{virieux1984sh, virieux1986p, moczo20023d, robertsson1994viscoelastic, cheng2021wave}, finite-element \citep{padovani1994low, koketsu2004finite}, or pseudo-spectral \citep{zhu2014modeling, wang2022propagating} methods, applied to the wave equation. While these approaches can, in principle, produce highly accurate results given a known subsurface velocity model, they are often prohibitively expensive for large-scale problems or when high-resolution solutions are required \citep{marfurt1984accuracy, wu2014optimized}. Moreover, in scenarios involving repeated evaluations, such as inversion loops or parameter studies, numerical simulations quickly become computational bottlenecks. 

Advances in machine learning have prompted various approaches to tackling forward modeling tasks governed by partial differential equations (PDEs), including the neural representation of seismic wavefields \citep{alkhalifah2021wavefield}. Physics-informed neural networks (PINNs) have emerged as one promising avenue \citep{raissi2019physics}. By embedding PDE constraints directly into their loss functions, PINNs can learn continuous solutions that respect the underlying physical laws, reducing the need for large labeled datasets. \cite{alkhalifah2020wavefield} pioneered the use of PINNs to solve the Helmholtz equation, enabling the neural representation of scattered wavefield solutions in isotropic media. \cite{song2021solving} further extended this approach to transversely isotropic media with a vertical symmetry axis. \cite{huang2022pinnup} introduced a novel training method called frequency upscaling, which enhances the ability of PINNs to represent high-frequency wavefields by pre-training a PINN on low-frequency wavefields and subsequently expanding the network's capacity through a neuron-splitting technique. Although PINNs have achieved some success in modeling wavefields, they often face challenges in scalability and efficiency \citep{cheng2024meta}. Training can become expensive as problem complexity grows, and their generalization to new velocity models and frequencies is limited, often requiring a full retraining for each new scenario. 

To address these limitations, recent developments in operator learning offer an alternative and potentially more powerful framework. Neural operators aim to learn mappings between function spaces rather than fixed-dimensional inputs, enabling them to handle entire families of PDE solutions \citep{li2020fourier}, such as a wide range of velocity models, without the need for retraining from scratch. This property makes neural operators particularly attractive for seismic applications. Instead of constructing a separate surrogate for each distinct velocity model or frequency, a trained operator can swiftly infer the corresponding wavefield \citep{huang2024}. By efficiently capturing the complex input-output relationships at the functional level, operator learning provides a scalable, flexible, and robust solution that can outperform PINNs in wavefield representations. \cite{wei2022small} developed a physics-informed fourier neural operator (FNO) method to approximate the seismic wavefield in the time domain. \cite{zhang2023learning} modified the vanilla FNO with enhanced Fourier kernel operations and layer connections to solve the 2D isotropic elastic wave equation, achieving 100-fold speedup over finite-difference methods and improving low-frequency predictions. \cite{li2023solving} introduced a paralleled FNO framework to efficiently train FNO-based solvers across variable velocity models and multiple source locations and, thus, to present the frequency domain seismic wavefields. In their implementation, the embedding of the source location is achieved by adding it as an input to the model. This input uses a one-hot encoding scheme across the entire spatial domain, where the source location is set to 1, while all other locations are set to 0. The frequency embedding, on the other hand, is represented by introducing a quantity of the same size as the model, with all values set to the frequency. Compared to their approach, \cite{huang2024} adopted a three-channel input, which not only includes the velocity model but also incorporates the background wavefield (real and imaginary parts). The background wavefield implicitly provides information about both the source location and the frequency. \cite{zou2024deep} proposed a 3-D neural Helmholtz operator for frequency-domain elastic wave modeling and, thus, address memory limitations of 3-D time-domain simulations. \cite{lehmann20243d} used the factorized FNO as an efficient and accurate surrogate model for 3-D elastic wave propagation. While neural operators exhibit remarkable flexibility and efficiency in representing seismic wavefield, many existing approaches still face challenges in achieving robust generalization to complex or out-of-distribution inputs. Real-world velocity models often feature heterogeneous and complicated structures that test the limits of current operator frameworks. As the complexity of the velocity model, conventional neural operators may suffer from reduced accuracy, stability issues, or the inability to capture subtle variations in wavefield patterns. 

To address these challenges, we propose a generative neural operator (GNO) that leverages the strengths of generative diffusion models (GDMs). GDMs have gained attention in various domains for their remarkable capacity to model complex data distributions through an iterative denoising processes \citep{ho2020denoising}. By using a GDM framework as a neural operator, we aim to build a model that not only learns a versatile forward mapping from velocity models and background wavefields to the scattered wavefields, but also exhibits enhanced generalization and stability. To further ensure the physical consistency of the generated wavefields, we incorporate a physics-guided sampling process into our GNO framework. Specifically, at each sampling step, we calculate a physical loss induced by the wave equation on intermediate results, and use its gradient to guide the reverse process. By guiding the wavefield sampling process towards physically consistent solutions, this strategy ensures that the generated scattered wavefields not only respect the input conditions but also adhere to the underlying physical laws of wave propagation. As a result, our proposed approach provides a powerful and scalable tool for representing multi-frequency seismic scattered wavefields across a broad range of subsurface scenarios.

In summary, our contributions are threefold:
\begin{itemize}
    \item We propose a generative neural operator (GNO) that integrates GDMs for the representation of seismic scattered wavefields, conditioned on the velocity model and background wavefield.
    \item We introduce a physics-guided sampling process, where physical loss induced by the wave equation is used to guide the generation process, ensuring the physical consistency of the generated wavefields.
    \item Our approach achieves enhanced generalization, stability, and accuracy across complex and heterogeneous velocity models, providing a scalable and efficient solution for multi-frequency seismic wavefield representation. We test the approach and its generalizablity on relatively complex velocity models.
\end{itemize}

\section{\textbf{Preliminaries: Generative diffusion models}}
Generative models aim to capture the underlying data distribution and generate new samples that are statistically consistent with the data distribution. Among various generative models, generative diffusion models (GDMs) have recently gained significant attention due to their ability to model complex, high-dimensional data distributions \citep{ho2020denoising}. 

GDMs are based on an iterative denoising process, where data is progressively corrupted with Gaussian noise in a forward process and subsequently reconstructed in a reverse process. The key idea is to model the data distribution through this noise-based diffusion mechanism. 

In the forward process, at each time step $t$, Gaussian noise is added to the data $x_0$, gradually transforming it into exactly a Gaussian noise at time step $T$. The process can be defined as a Markov chain:
\begin{equation}\label{eq1}
    q(x_t | x_{t-1}) = \mathcal{N}(x_t; \sqrt{\alpha_t} x_{t-1}, (1 - \alpha_t) \mathbf{I}),
\end{equation}
where $x_t$ is the noisy data at step $t$, $\alpha_t = 1 - \beta_t$ defines the rate of noise addition, $\beta_t$ is the noise schedule, the identity matrix $\mathbf{I}$ denotes the isotropic Gaussian noise, and $\mathcal{N}$ denotes a Gaussian distribution. 

The reverse process aims to recover the clean data $x_0$ starting from a sample drawn fro a normal distribution, $x_T \sim \mathcal{N}(0,\mathbf{I})$. This is achieved by learning a parameterized model $p_\theta$ that predicts the intermediate data distribution at each step $t$:
\begin{equation}\label{eq2}
    p_\theta(x_{t-1} | x_t) = \mathcal{N}(x_{t-1}; \mu_\theta(x_t, t), \Sigma_\theta(x_t, t)),
\end{equation}
where $\mu_\theta(x_t, t)$ and $\Sigma_\theta(x_t, t)$ are the mean and the variance, which can be expressed by the trained parameterized model $p_\theta$, and $x_{t-1}$ is the data at the previous step. Here, the variance introduces randomness in the reverse process and, thus, improve the diversity of the generated images. Commonly, $\Sigma_\theta(x_t, t)$ is set to a fixed schedule such as $\Sigma_\theta(x_t, t) = \sigma_t^2\mathbf{I}$, where $\sigma_t$ controls the level of the randomness. 

To compute the mean $\mu_\theta$, an NN is trained to predict the noise $\epsilon$ added at each step. Given the noisy data $x_t$ and time step $t$, the predicted mean can be expressed as:
\begin{equation}\label{eq3}
    \mu_\theta(x_t, t) = \frac{1}{\sqrt{\alpha_t}} \left( x_t - \frac{1 - \alpha_t}{\sqrt{1 - \bar{\alpha}_t}} \epsilon_\theta(x_t, t) \right),
\end{equation}
where $\epsilon_\theta(x_t, t)$ is the NN's prediction of the noise at step $t$, and $ \bar{\alpha}_t = \prod_{s=1}^t \alpha_s$ represents the cumulative product of noise schedules up to step $t$.

To train the network $\epsilon_\theta$, we can minimize the mean squared error (MSE) between the true noise $\epsilon \sim \mathcal{N}(0, \mathbf{I})$ added at the diffusion step $t$ and the predicted noise $\epsilon_\theta(x_t, t)$:
\begin{equation}\label{eq4}
    \mathcal{L}(\theta) = \mathbb{E}_{x_0, t, \epsilon} \left[ \| \epsilon - \epsilon_\theta(x_t, t) \|^2 \right].
\end{equation}
Once the network $\epsilon_\theta$ is trained, the reverse diffusion process can be used to generate clean data $x_0$ from pure Gaussian noise $x_T$. Starting from $x_T \sim \mathcal{N}(0, \mathbf{I})$, the data at step $t-1$ can be sampled as:
\begin{equation}\label{eq5}
    x_{t-1} = \mu_\theta(x_t, t) + \sigma_t z, \quad z \sim \mathcal{N}(0, \mathbf{I}).
\end{equation}

Therefore, we can see that through the forward diffusion process, clean data is progressively transformed into Gaussian noise. In the reverse process, an NN is trained to predict the noise at each step, allowing the computation of the mean $ \mu_\theta $ needed to reconstruct the clean data. During sampling, the reverse process iteratively removes the noise starting from $ x_T \sim \mathcal{N}(0, \mathbf{I}) $, ultimately generating data $ x_0 $ that matches the original data distribution. 

However, the DDPM sampling process is inherently stochastic due to the random noise introduced by the variance $\Sigma_\theta(x_t, t)$ at each step. While this randomness contributes to the diversity of generated samples, it also results in slower generation and less control over the generation process. To address these limitations, the denoising diffusion implicit model (DDIM) \citep{song2020denoising} modifies the reverse process by removing the stochasticity, enabling faster sampling and more precise control over the generation process. Instead of introducing randomness at each step, DDIM reformulates the reverse process as a non-Markovian transformation that directly maps noise $x_T$ to clean data $x_0$ in fewer number of steps, as follows:
\begin{equation}\label{eq6}
    x_{t-1} = \sqrt{\bar{\alpha}_{t-1}} x_0 + \sqrt{1 - \bar{\alpha}_{t-1}} \epsilon_\theta(x_t, t),
\end{equation}
where $x_0$ is explicitly predicted using the NN as:
\begin{equation}\label{eq:ddim_x0}
    x_0 = \frac{x_t - \sqrt{1 - \bar{\alpha}_t} \epsilon_\theta(x_t, t)}{\sqrt{\bar{\alpha}_t}}.
\end{equation}

By removing the reliance on random noise, DDIM achieves a deterministic sampling process while maintaining high-quality generation results. This deterministic property also makes DDIM particularly useful in tasks requiring precise and consistent outputs. Inspired by the success of DDIM in modeling complex distributions, we leverage this framework to represent seismic scattered wavefields. In the following sections, we will introduce our proposed generative neural operator (GNO), which incorporates the principles of GDMs to generate scattered wavefields while ensuring physical consistency.

\section{\textbf{Method}}
This section presents the methodology underlying our proposed GNO framework for seismic scattered wavefield representation. We begin by introducing the Helmholtz equation and the corresponding scattered wavefield version of it, which serves as the governing equation for constraining the solution. Next, we introduce our GNO, which leverages GDMs to generate scattered wavefields conditioned on the input velocity model and background wavefield. To ensure the physical consistency of the generated wavefields, we further incorporate a physics-guided sampling process, where physical loss gradients derived from the Helmholtz equation are used to guide intermediate sampling results. Finally, we describe the network architecture of our GNO, detailing the input-output design, key components, and implementation strategies.

\subsection{The Helmholtz equation}
The wave equation, which describes the wave propagation, is often solved in the time domain using numerical methods such as finite-difference, finite-element, and pseudo-spectral schemes. While these methods are highly accurate, they suffer from significant computational cost and memory requirements, especially for large-scale or high-resolution simulations. To overcome these limitations, the wave equation can be reformulated in the frequency domain as the Helmholtz equation, which allows for efficient frequency-by-frequency solutions, thus reducing dimensionality of the problem \citep{pratt1999seismicpart1}. 

The Helmholtz equation in an acoustic, isotropic, constant-density medium can be expressed as:
\begin{equation}\label{eq8}
    (\nabla^2 + k^2)u(\mathbf{x}) = f(\mathbf{x}), \quad \text{where} \quad k = \frac{\omega}{v}.
\end{equation}
Here, $u(\mathbf{x})$ represents the complex-valued wavefield, $\mathbf{x}$ is the spatial coordinate, $\omega$ is the angular frequency, $v$ is the velocity model, and $\nabla^2$ is the Laplacian operator with respect to the spatial coordinates. The wavefield $u(\mathbf{x})$ consists of both the real and imaginary parts, which together hold the amplitude and phase information of the wavefield. 

For a point source located at $\mathbf{x_s}$, the Helmholtz equation exhibits a singularity in the wavefield solution at the source, which is problematic for numerical methods. To address this singularity, we decompose the total wavefield $u$ into a background wavefield $u_0$ and a scattered wavefield $\delta u$:
\begin{equation}\label{eq9}
    \delta u(\mathbf{x}) = u(\mathbf{x}) - u_0(\mathbf{x}).
\end{equation}
The background wavefield $u_0(\mathbf{x})$ satisfies the Helmholtz equation for a background velocity $ v_0 $, and the scattered wavefield satisfies \citep{alkhalifah2021wavefield}:
\begin{equation}\label{eq10}
    \left(\nabla^2 + \frac{\omega^2}{v^2} \right)\delta u(\mathbf{x}) = -\omega^2 \delta m u_0(\mathbf{x}), \quad \text{where} \quad \delta m = \frac{1}{v^2} - \frac{1}{v_0^2}.
\end{equation}

In 2D isotropic media, for a constant background velocity $v_0$, we can use an analytical equation to calculate the background wavefield $u_0(\mathbf{x})$ as follows \citep{Aki1980QuantitativeST}:
\begin{equation}\label{eq11}
    u_0(\mathbf{x}) = \frac{\text{i}}{4} H_0^{(2)}\left(\frac{\omega}{v_0} |\mathbf{x} - \mathbf{x_s}|\right),
\end{equation}
where $H_0^{(2)}$ is the zero-order Hankel function of the second kind, $\text{i}$ is the imaginary unit, and $\left| \cdot \right|$ represents the Euclidean distance. 

Compared with using numerical methods to solve the background wavefield, the analytical equation provides a more efficient way to quickly compute background wavefield solutions. On the other hand, the use of an analytical solution for the background wavefield provides significant convenience, as it implicitly encodes both the frequency and source location information of the wavefield \citep{huang2024}. When we use GDMs to model scattered wavefield, the frequency and source location information embedded in the background wavefield serve as critical constraints to control the generation process. In the next section, we will introduce how our proposed GNO, which is motivated from GDMs, leverages the background wavefield and velocity model to generate scattered wavefields.

\subsection{Generative neural operator for representing scattered wavefield}
\begin{figure}[htbp]
\centering
\includegraphics[width=0.88\textwidth]{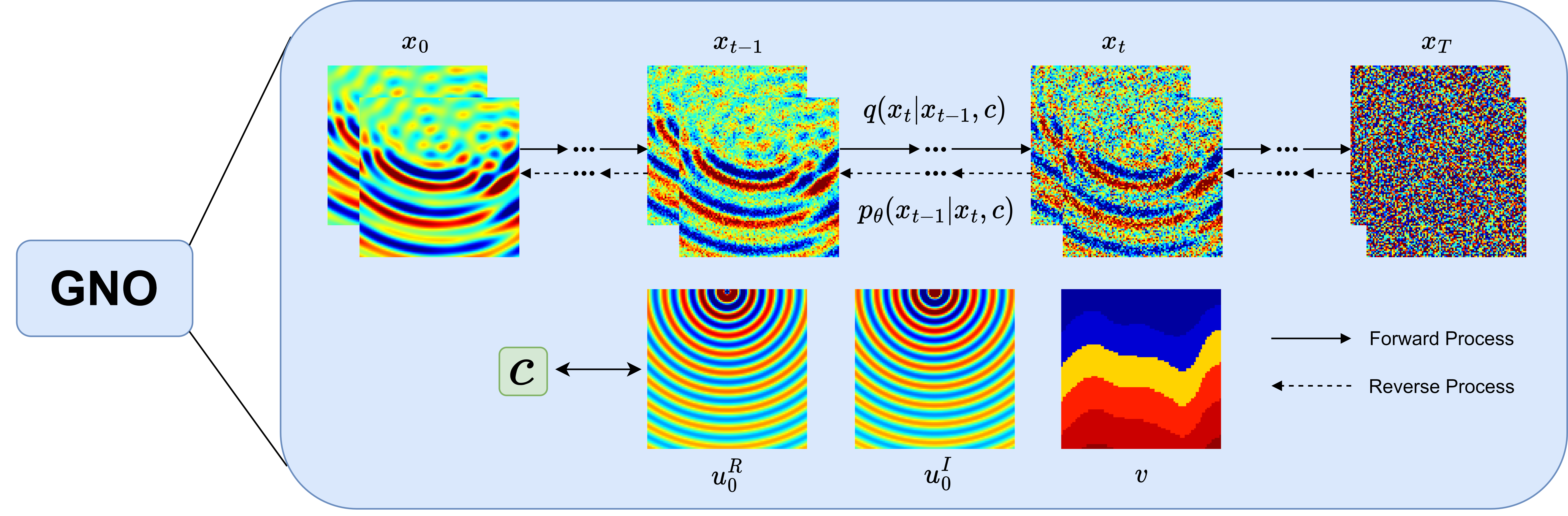}
\caption{An illustration of our GNO framework. In the forward process ($q(x_t | x_{t-1}, c)$, solid arrows), the clean scattered wavefield $x_0$ is gradually corrupted by Gaussian noise across diffusion steps, transforming into Gaussian noise $x_T$ at the final step $T$. The reverse process ($p_\theta(x_{t-1} | x_t, c)$, dashed arrows) aims to reconstruct a clean wavefield $x_0$ from Gaussian noise $x_T$ by iteratively denoising intermediate samples $x_t$. The condition input is $c = (u_0^R, u_0^I, v)$, which includes the real and imaginary parts of the background wavefield ($u_0^R, u_0^I$) and the velocity model ($v$).}
\label{fig1}
\end{figure}

Our proposed GNO framework, illustrated in Figure \ref{fig1}, inherits the forward-reverse diffusion process from GDMs. Specifically, let the scattered wavefield in the frequency domain be denoted as $x_0$, including its real and imaginary parts. The forward diffusion process is defined as a gradual transition from the scattered wavefield distribution to a Gaussian distribution. For given diffusion steps $t = 0, \ldots, T$, this forward process can be expressed as in Equation \ref{eq1}. At $t = T$, the scattered wavefield becomes exactly a Gaussian noise, denoted by $x_T$. 

In the reverse process, an NN is trained to denoise the scattered wavefield at various noise levels. During training, the forward process is used to map the original scattered wavefield $x_0$ into noisy samples $x_t$ at different diffusion steps, thereby constructing training samples. The network input, handled as five channels, consists of: 
\begin{enumerate}
    \item The noisy scattered wavefield $x_t$, including both real and imaginary parts (2 channels).
    \item The conditional input $c = (u_0^R, u_0^I, v)$, where $u_0^R$ and $u_0^I$ represent the real and imaginary parts of the background wavefield, and $v$ is the velocity model (3 channels).
\end{enumerate}
The background wavefield implicitly provides information about the source location and the frequency of the wavefield, thereby eliminating the need to explicitly encode these features. 

Unlike conventional GDMs, where the network is trained to predict the noise added at each diffusion step, our approach trains the network to directly predict the scattered wavefield $x_0$. This design enables the GNO to learn how to reconstruct the noise-free scattered wavefield $x_0$ from any noisy state $x_t$. By explicitly targeting $x_0$, the network achieves a more direct and efficient learning process. So, the output of the network is two channels given by the real and imaginary parts of the denoised scattered wavefield. Alternatively, we could have trained the network to predict the full wavefield directly. However, we wanted the network to focus on (and represent the distribution of) the lower amplitude scattered part of the wavefield, and the full wavefield can be evaluated by simply adding the analytical computed background wavefield. 

Once the model is trained to predict the clean scattered wavefield $x_0$ from a noisy input $x_t$, we employ DDIM \citep{song2020denoising} to further accelerate the sampling process. Notably, the generation process remains conditioned on the background wavefield $(u_0^R, u_0^I)$ and the velocity model $v$, ensuring that the synthesized wavefields meet specified requirements, such as source location, frequency content, and subsurface parameters. For each diffusion step $t$, the network predicts the target scattered wavefield $x_{0,\theta}(x_t, c, t)$ given the conditions $c$. Using DDIM, the reverse update from $x_t$ to $x_{t-1}$ can then be expressed as:
\begin{equation}\label{eq12}
x_{t-1} = \sqrt{\bar{\alpha}_{t-1}} \, x_{0,\theta}(x_t, c, t) 
\;+\; \sqrt{1 - \bar{\alpha}_{t-1} } \,\hat{\epsilon}(x_t, c, t),
\end{equation}
where the estimated noise $\hat{\epsilon}(x_t, c, t)$ at step $t$ is computed as:
\begin{equation}\label{eq13}
\hat{\epsilon}(x_t, c, t) = \frac{x_t - \sqrt{\bar{\alpha}_t}\, x_{0,\theta}(x_t, c, t)}{\sqrt{1 - \bar{\alpha}_t}}.
\end{equation}

By leveraging the non-Markovian nature of DDIM and the network’s direct predictions of $x_0$ under specified conditions, the number of sampling steps can be effectively reduced from $T$ to a smaller number $L$, where $L \ll T$. This acceleration significantly enhances the computational efficiency of the generation process while maintaining high-quality scattered wavefield representations. This efficiency is critical for applications such as seismic inversion, where rapid wavefield generation is essential for iterative updates.

As we know, GDMs are commonly used in computer vision tasks to synthesize diverse images. In contrast, we, here, present a conditional GDM framework tailored for seismic wavefield solutions. The primary goal is not to generate a large variety of wavefields but to produce accurate wavefields that correspond to specific velocity models, frequencies, and source locations. Instead of focusing on diversity, we leverage the powerful capability of GDMs to model complex data distributions, ensuring that the generated wavefields are precise and meet our strict requirements. This precision is crucial for seismic inversion, where accurate wavefield representations are more important than diversity. 

\subsection{Physics-guided wavefield generation}
To enhance the physical consistency of the generated scattered wavefields, we introduce a physics-guided sampling process that leverages the governing physical equation, i.e., the scattered wavefield equation (Equation \ref{eq10}). In this process, each intermediate sampling result at every diffusion step is refined using gradients derived from the PDE loss. This ensures that the generated wavefields are not only conditioned on the input velocity model and background wavefield, but also satisfy the underlying physical laws. 

At each diffusion step $t$, our objective is for the intermediate wavefield $x_t$ to satisfy the scattered wavefield equation as closely as possible. Mathematically, this means that the PDE residual:
\begin{equation}\label{eq14}
    \mathbf{r} = \left( \nabla^2 + \frac{\omega^2}{v^2} \right) x_t + \omega^2 \delta m u_0
\end{equation}
should approach zero at every sampling step. In other words, we aim to minimize the inconsistency between the intermediate sampling product and the scattered wavefield equation. To achieve this, we define a PDE loss function $\mathcal{L}_{\text{PDE}}$ that quantifies this residual $\mathbf{r}$:
\begin{equation}\label{eq15}
    \mathcal{L}_{\text{PDE}}(x_t) = \left\| \mathbf{r} \right\|_2^2 .
\end{equation}
In practice, the Laplacian operator $\nabla^2$ in the loss function is approximated using the finite difference method on a discrete spatial grid. As a result, we can numerically compute the residual at each grid point. 

To correct the intermediate wavefield $x_t$, we then compute the gradient of the PDE loss with respect to $x_t$:
\begin{equation}\label{eq16}
    \nabla_{x_t} \mathcal{L}_{\text{PDE}} = \frac{\partial \mathcal{L}_{\text{PDE}}(x_t)}{\partial x_t}= 2\cdot\mathbf{r}\cdot \nabla_{x_t}\mathbf{r}.
\end{equation}
This gradient provides the direction in which the intermediate wavefield $x_t$ should be adjusted to better satisfy the scattered wavefield equation. The corrected wavefield $\tilde{x}_t$ is then obtained as:
\begin{equation}\label{eq17}
    \tilde{x}_t = x_t - \eta \cdot \nabla_{x_t} \mathcal{L}_{\text{PDE}},
\end{equation}
where $\eta$ is a scaling factor that controls the magnitude of the correction, like the learning rate. By iteratively applying this correction at each diffusion step, the sampled wavefield is guided towards a physically consistent solution. 

The physics-guided correction process is seamlessly integrated into DDIM framework. Specifically, at each step $t$, we first update the intermediate wavefield $x_t$ using the DDIM sampling rule (Equation \ref{eq12}), followed by the physics-guided correction (Equation \ref{eq17}). This combination ensures that the generated wavefields not only match the condition inputs, such as the velocity model and background wavefield, but also respect the physical laws governing wave propagation. 

By incorporating this physics-guided correction process, our framework effectively generates wavefields that are not only consistent with the data distribution modeled by the generative diffusion process but also satisfy the physical constraints imposed by the scattered wavefield equation. This dual consistency significantly enhances the reliability and interpretability of the generated wavefields. 

\subsection{Network architecture}
\begin{figure}[!htb]
\centering
\includegraphics[width=0.95\textwidth]{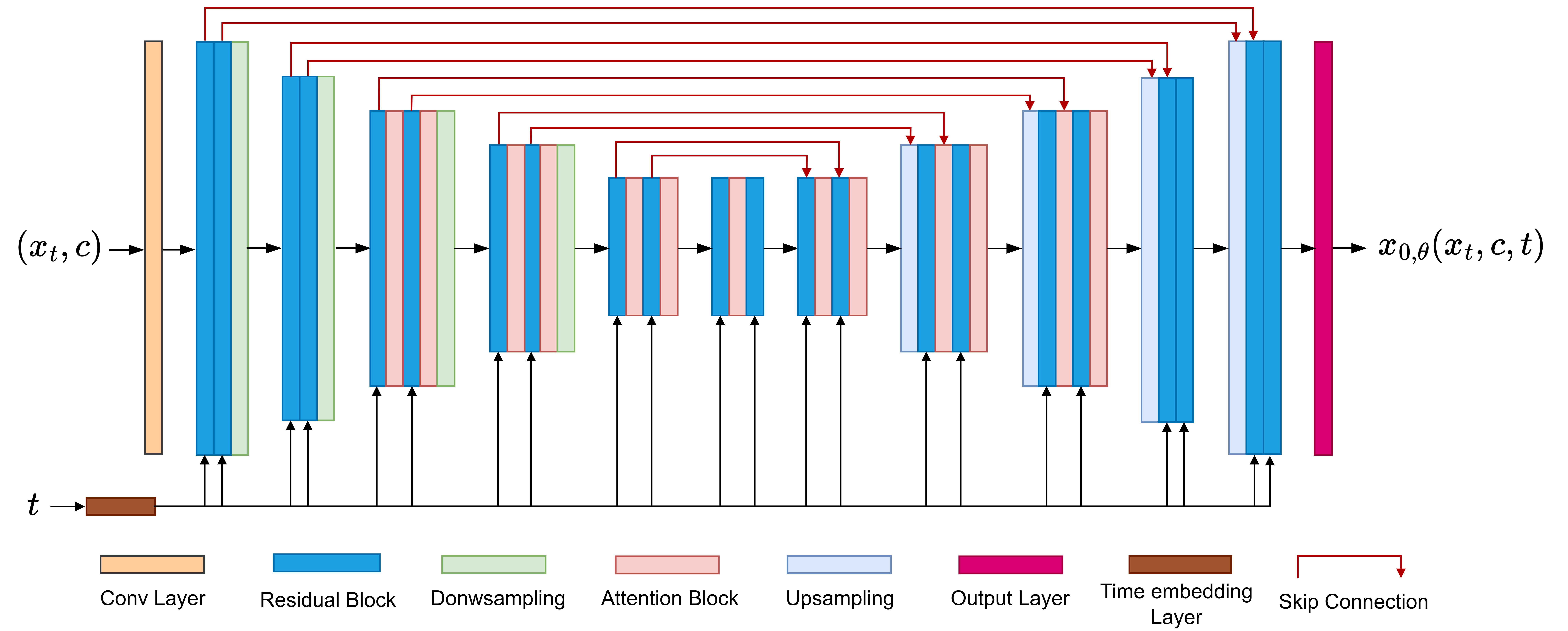}
\caption{An illustration of our network architecture. The input to the network is the noisy scattered wavefield, $x_t$ and the conditions, $c$, and the output is the denoised wavefield, $x_0$, at time step, $t$.}
\label{fig2}
\end{figure}

Our GNO employs a U-Net-based architecture, as illustrated in Figure \ref{fig2}. This architecture is designed to predict the scattered wavefield $x_{0,\theta}(x_t, c, t)$ conditioned on the noisy wavefield $x_t$, background wavefield $u_0^R, u_0^I$, velocity model $v$, and diffusion time step $t$. The key components and workflow of the architecture are described as follows:

\begin{enumerate}
    \item \textbf{Input and conditioning:}  
    As we stated before, the input to the network consists of: 1. $x_t$: The noisy scattered wavefield at diffusion step $t$; 2. $c = (u_0^R, u_0^I, v)$: The conditional input, including the background wavefield's real and imaginary parts and the velocity model. The input data is first processed by an initial 3$\times$3 convolution layer to extract low-level features.
    
    \item \textbf{Time embedding layer:}  
    The diffusion time step $t$ is processed through a time embedding layer, encoding temporal information into a feature vector. This embedding is added to the residual blocks in both the encoder and decoder paths to ensure time-specific information is effectively integrated.
    
    \item \textbf{Encoder path:}  
    The encoder consists of multiple stages, each stage includes residual blocks and a downsampling layer, where the downsampling layer includes a convolution layer with a stride of 2. Notably, attention blocks are selectively applied at certain stages of the encoder to enhance long-range dependencies and emphasize key features, rather than being uniformly present at every stage.

    \item \textbf{Decoder path:}  
    The decoder mirrors the encoder with multiple stages. Also, each stage consists of residual blocks and an upsampling layer, where the upsampling layer first performs a nearest interpolation to double the spatial resolution of the feature maps, followed by a 3$\times$3 convolution layer. Similar to the encoder, attention blocks are selectively included at certain stages of the decoder to focus on important features and improve feature representation.

    \item \textbf{Skip connections:}  
    Each encoder layer is connected to its corresponding decoder layer through skip connections (red arrows in Figure \ref{fig2}). These connections allow low-level spatial information to be directly passed to the decoder, enhancing the reconstruction of fine details in the predicted wavefield.

    \item \textbf{Output layer:} 
    The final feature maps are passed through a group normalization layer, followed by a sigmoid linear unit function, and finally refined using a 3$\times$3 convolution layer to produce the clean scattered wavefield $x_{0,\theta}(x_t, c, t)$. 
\end{enumerate}
Here, the structure of some key components, including residual blocks, attention blocks, and time embedding layers, are detailed in \cite{cheng2025gsfm}.

\section{\textbf{Numerical examples}}
In the following, we will evaluate our proposed GNO. First, we describe the training dataset and details. Second, we show the impact of the physics-guided sampling strategy on the generated wavefields. Third, we demonstrate the model’s performance on test cases that match the training data distribution (in-distribution tests). Finally, we assess the model’s ability to generalize by applying it to velocity models that were not seen during training (out-of-distribution tests). 

\subsection{Data and training}
To prepare for the training of our GNO, we construct a training dataset derived from the OpenFWI, a popular large-scale multi-structural benchmark datasets \citep{deng2022openfwi}. This dataset contains multiple classes, such as FlatVel-A, FlatVel-B, CurveVel-A, CurveVel-B, FlatFault-A, FlatFault-B, CurveFault-A, and CurveFault-B. Each class of velocity models represents different geological features and velocity distributions. For our training dataset, we only select a limited 2000 velocity models of size $70\times70$ from the CurvevVelA class, and then resize each to a resolution of $101\times101$. Using the selected velocity models, we generate four smoothed versions of each using Gaussian smoothing, where the smoothing parameter $\sigma$ randomly varies between 2 and 25. These augmentations increase the samples to 10000 velocity models. For each velocity model, the dataset includes seven discrete frequencies from 4 Hz to 10 Hz. Each velocity-frequency pair then uses five different source positions, resulting in approximately 350000 training samples. We do this because we hope the GNO learns diverse conditions, thereby ensuring robust generalization and efficient wavefield generation and, thus, effectively serving potentially future inversion tasks. Each sample includes both the real and imaginary parts of the scattered and background wavefields, as well as the corresponding velocity model. The scattered wavefield is generated using a finite-difference solver, while the background wavefield is calculated using an analytical formulation \ref{eq11}. 

Our GNO training uses a fixed learning rate at 1e-4, and a batch size of 64. We use an AdamW optimizer \citep{loshchilov2017decoupled}, with an exponential moving average and a decay rate of 0.999 to stabilize the training process. The GNO is trained for 200000 iterations on a single NVIDIA A100 (80 GB) GPU, taking about 28 hours to complete. For comparison, we also train a Fourier neural operator (FNO) on the same dataset as our benchmark \citep{huang2024}. The FNO takes in three input channels: the real and imaginary parts of the background wavefield, and the velocity model. The target output for the FNO is the scattered wavefield (real and imaginary parts). The FNO consists of 4 FNO layers, where each layer is configured with a truncated Fourier mode of 48 and width of 128. We set the batch size to 64, and train for 200000 iterations as well. The initial learning rate is $1\times10^{-3}$ and is decayed by a factor of 0.97 every 100 iterations. Under these settings, the FNO requires approximately 24 hours of training time. 

\subsection{The effects of physics-guide sampling}
\begin{figure}[htbp]
\centering
\includegraphics[width=0.4\textwidth]{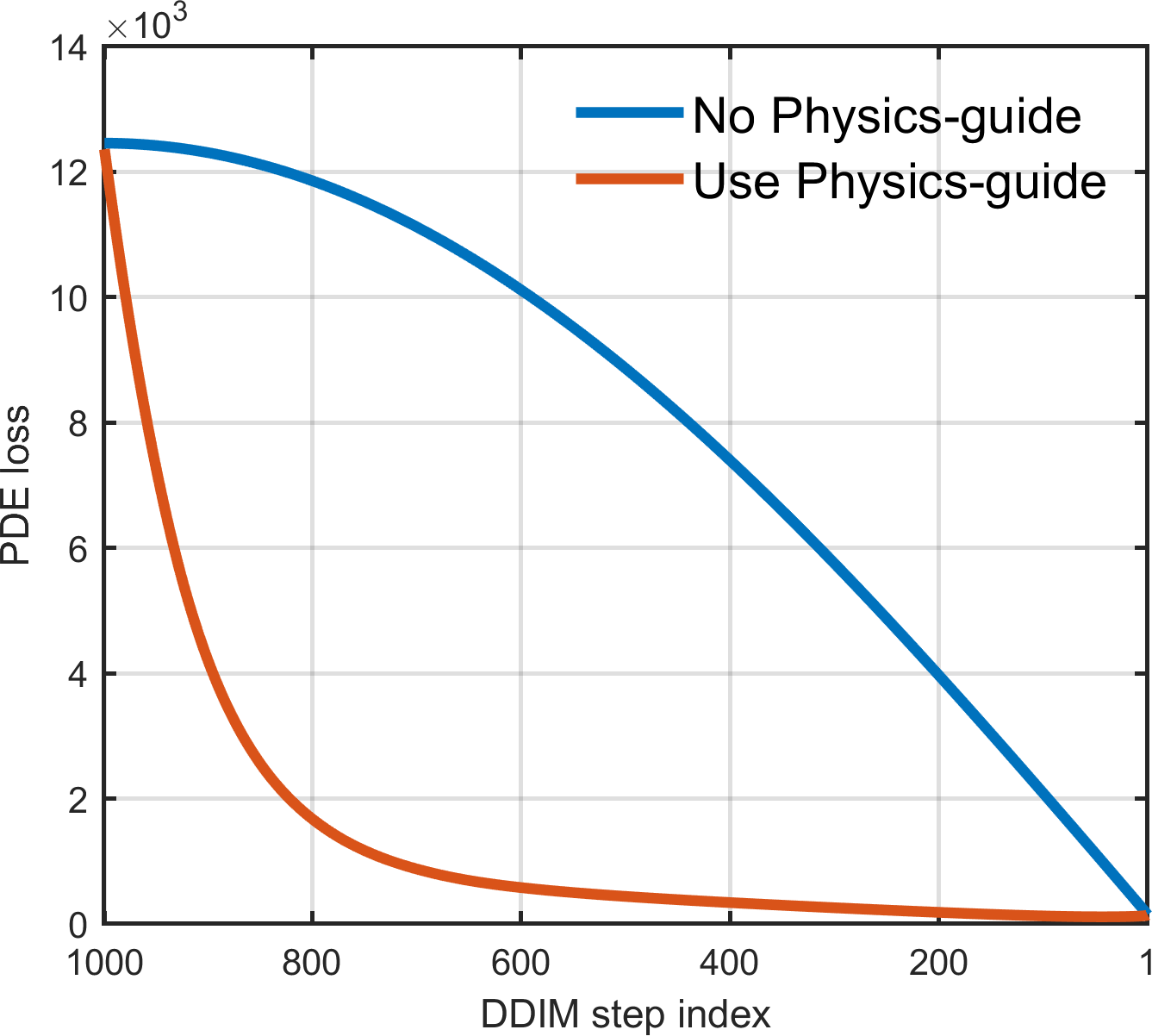}
\caption{Comparison of PDE loss during the DDIM sampling process for cases without (blue) and with (orange) physics guidance.}
\label{fig3}
\end{figure}

We first demonstrate how our physics-guided sampling strategy improves the quality and convergence speed of the generated wavefields. Figure~\ref{fig3} displays the PDE loss with respect to the DDIM step index for cases with and without physics guidance. We can see that, when physics-guided sampling is used (orange curve), the PDE loss rapidly decreases and remains below that of the case without physics guidance (blue curve) throughout the sampling process. This indicates that incorporating physical constraints at each sampling step effectively drives the intermediate wavefields toward physically consistent solutions. 

\begin{figure}[htbp]
\centering
\includegraphics[width=0.95\textwidth]{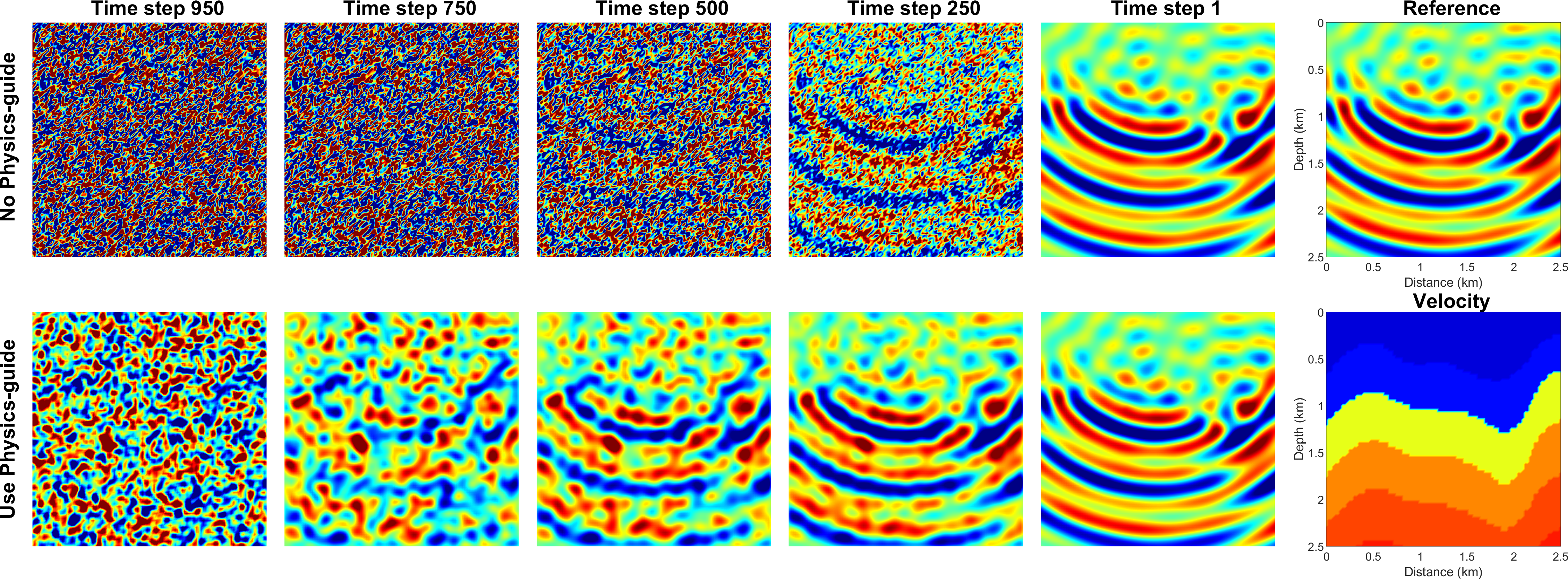}
\caption{Comparison of intermediate scattered wavefields generated without (top row) and with (bottom row) physics guidance at DDIM steps 950, 750, 500, 250, and 1 (left to right). The topmost figure in the last column shows the numerically simulated reference wavefield, and the bottom figure in the last column shows the corresponding velocity model. The source is located at in the center of the top surface, and the frequency is 6 Hz.}
\label{fig4}
\end{figure}

Figure~\ref{fig4} illustrates the evolution of intermediate wavefields at different sampling steps for two scenarios: without physics guidance (top row) and with physics guidance (bottom row). The five columns from left to right correspond to time steps 950, 750, 500, 250, and 1, respectively. The last column on the right shows the reference wavefield obtained by a numerical solver (top) and the corresponding velocity model (bottom), both belonging to the same CurveVel-A class as in the training set. It is evident that, with physics guidance, the network quickly captures the main wavefield structures, whereas without guidance, the generated intermediate wavefield still contains considerable noise even at step 250. 

\begin{figure}[htbp]
\centering
\includegraphics[width=0.4\textwidth]{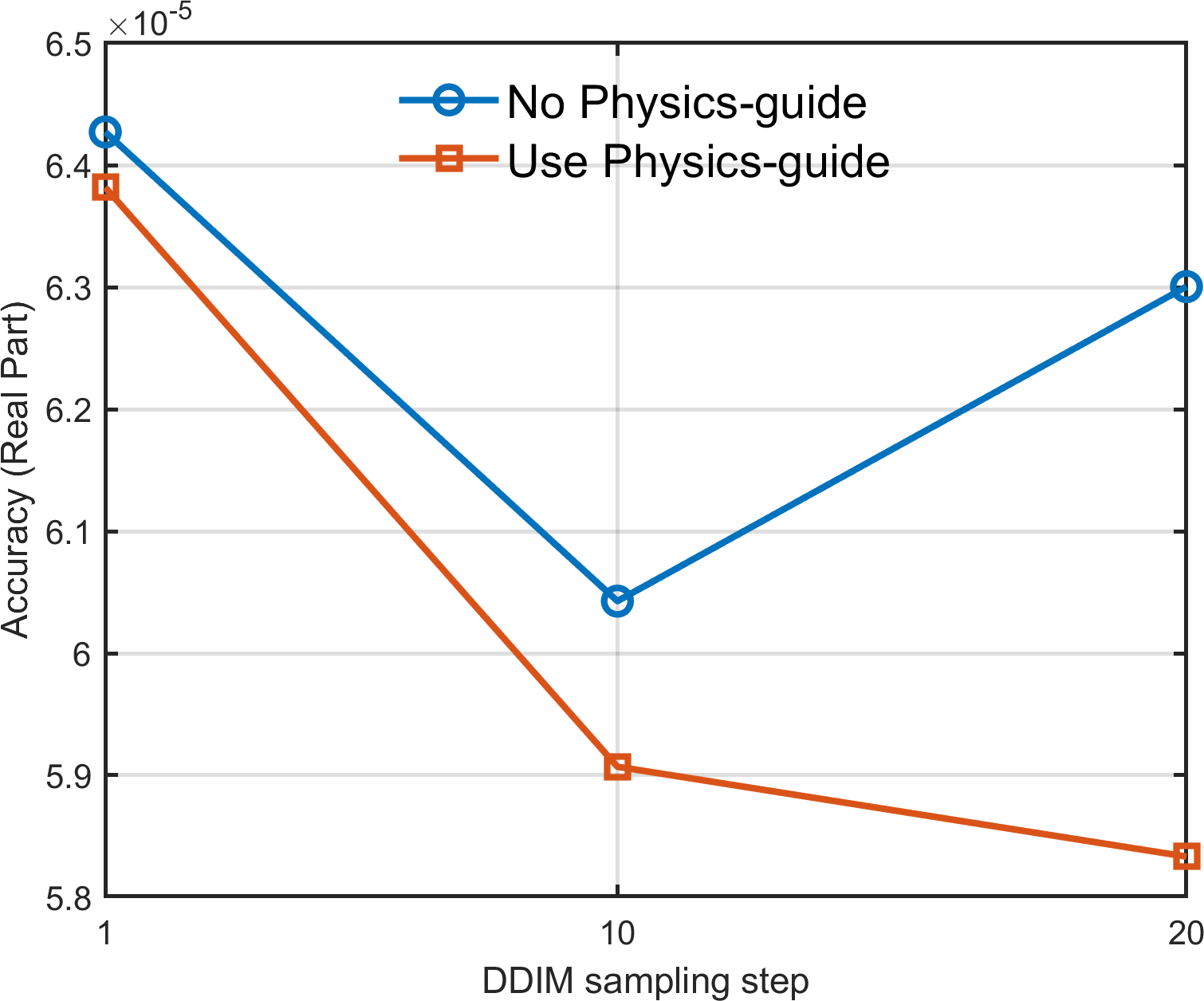}
\caption{Accuracy of the generated real-part scattered wavefields, measured by mean squared error relative to the reference solution, where we reduce the DDIM sampling steps to 1, 10, and 20.}
\label{fig5}
\end{figure}

We further compare the accuracy of the real-part scattered wavefield when reducing the number of DDIM sampling steps to 1, 10, and 20 (see Figure~\ref{fig5}). The accuracy metric is computed as the MSE between the generated wavefield and the reference wavefield. Across all three step settings, the presence of physics guidance (orange curve) yields noticeably lower MSE than the case without guidance (blue curve). This difference is especially significant when using 20 sampling steps, confirming that physics-guided sampling substantially enhances our GNO’s ability to represent the wavefield accurately. 

\begin{figure}[htbp]
\centering
\includegraphics[width=0.95\textwidth]{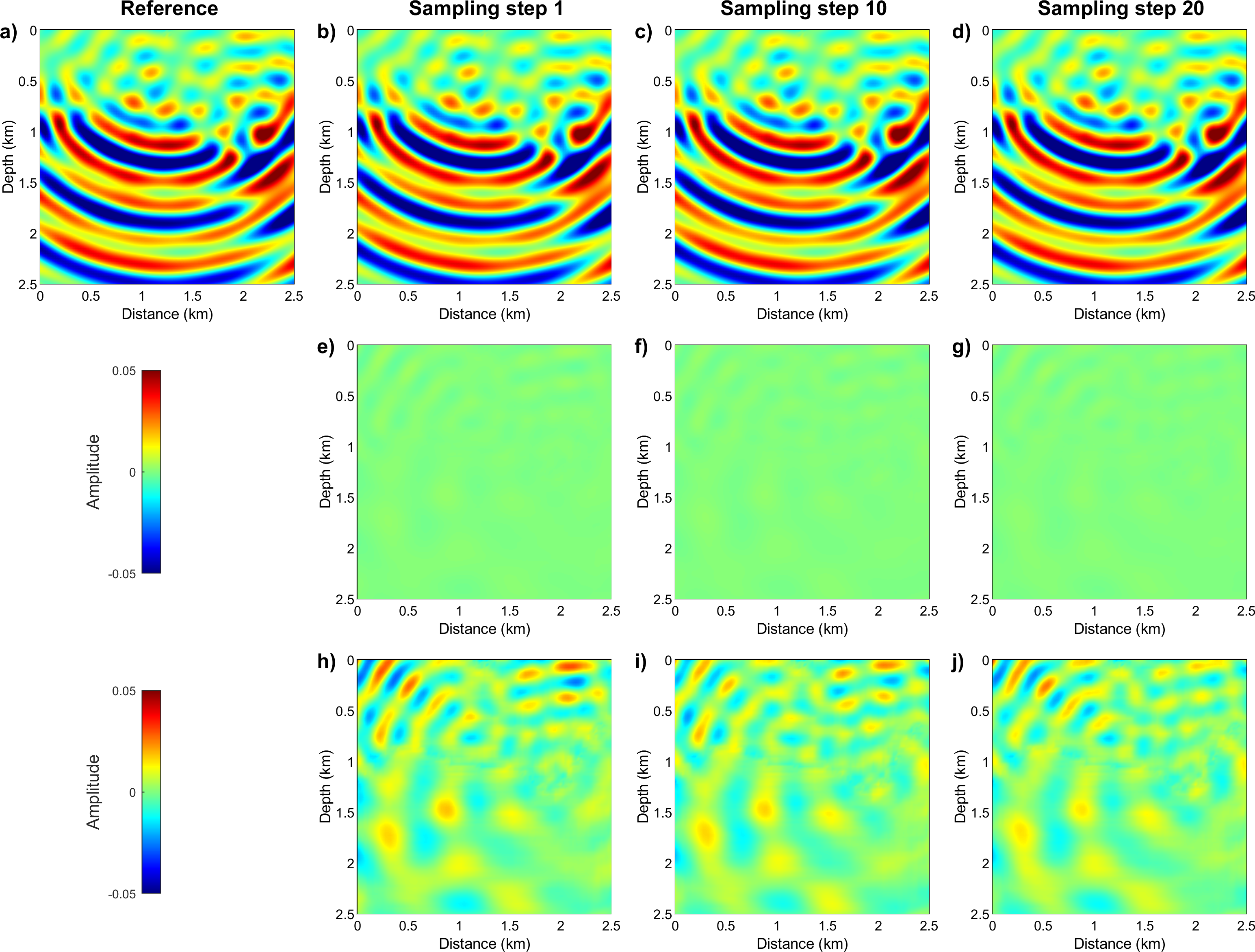}
\caption{Scattered wavefields (real part) generated with physics guidance using 1, 10, and 20 DDIM steps. The top row shows the reference wavefield (left) and the generated wavefields at steps 1, 10, and 20. The middle row displays the corresponding residuals at the original scale, and the bottom row shows those residuals magnified by a factor of 10 to highlight subtle differences.}
\label{fig6}
\end{figure}

Figure~\ref{fig6} shows examples of real-part scattered wavefields (real part) generated with physics-guided sampling for 1, 10, and 20 DDIM steps. The top row presents (from left to right) the reference wavefield and the wavefields generated at steps 1, 10, and 20, the middle row displays their corresponding residuals with respect to the reference, while the bottom row show the same residuals but magnified by a factor of 10. We can see that, although all three generated wavefields appear visually similar at normal scale, the magnified residuals reveal that the single-step solution exhibits noticeably larger errors than those generated using 10 or 20 steps. This observation suggests that allowing more sampling iterations refines the wavefield representation. Nonetheless, to maintain a favorable balance between accuracy and computational efficiency, we adopt a single-step generation in subsequent experiments. 

\subsection{Test on in-distribution scenarios}
\begin{figure}[htbp]
\centering
\includegraphics[width=0.95\textwidth]{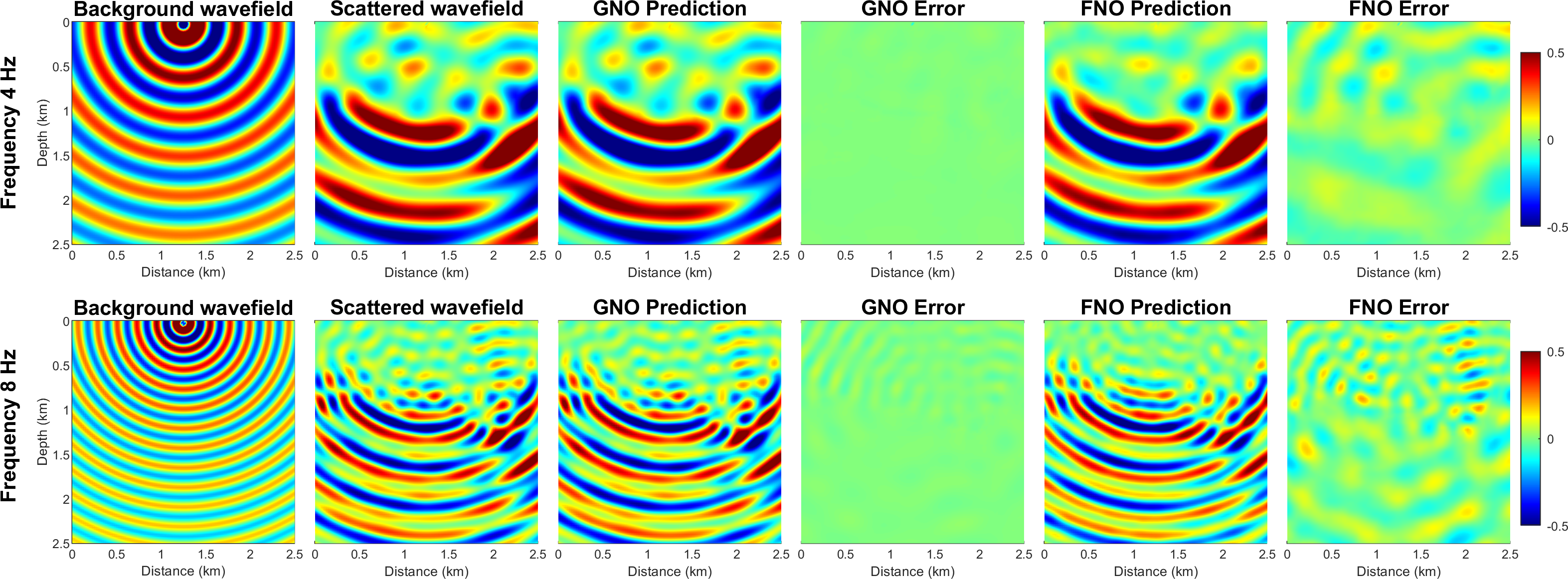}
\caption{Comparison of real-part scattered wavefield predictions for an in-distribution velocity model at 4 Hz (top row) and 8 Hz (bottom row). From left to right: background wavefield (real part), reference scattered wavefield (real part), GNO-predicted scattered wavefield (real part), GNO residual (reference minus GNO) plotted at the same scale, FNO-predicted scattered wavefield (real part), and FNO residual (reference minus FNO) plotted at the same scale.}
\label{fig7}
\end{figure}

We now evaluate our GNO on in-distribution test.
We use the same velocity model from the previous section, which belongs to the CurveVel-A class used during training. Here, we focus only in our illustrations on the real part of the scattered wavefield to avoid redundancy, omitting the imaginary part as they provide similar results. Figure \ref{fig7} compares our GNO-generated wavefields and those of the FNO at two representative frequencies, 4Hz and 8Hz (both of which fall within the training range). For each row, from left to right, we display the background wavefield, the reference scattered wavefield, the GNO-predicted scattered wavefield, the residual between GNO and the reference, the FNO-predicted scattered wavefield, and the residual between FNO and the reference. It is evident that our GNO solutions closely match the numerical reference wavefields, producing very small residuals for both frequencies. By contrast, the FNO results have visibly larger errors, especially at the higher frequency of 8 Hz. 

\begin{figure}[htbp]
\centering
\includegraphics[width=0.95\textwidth]{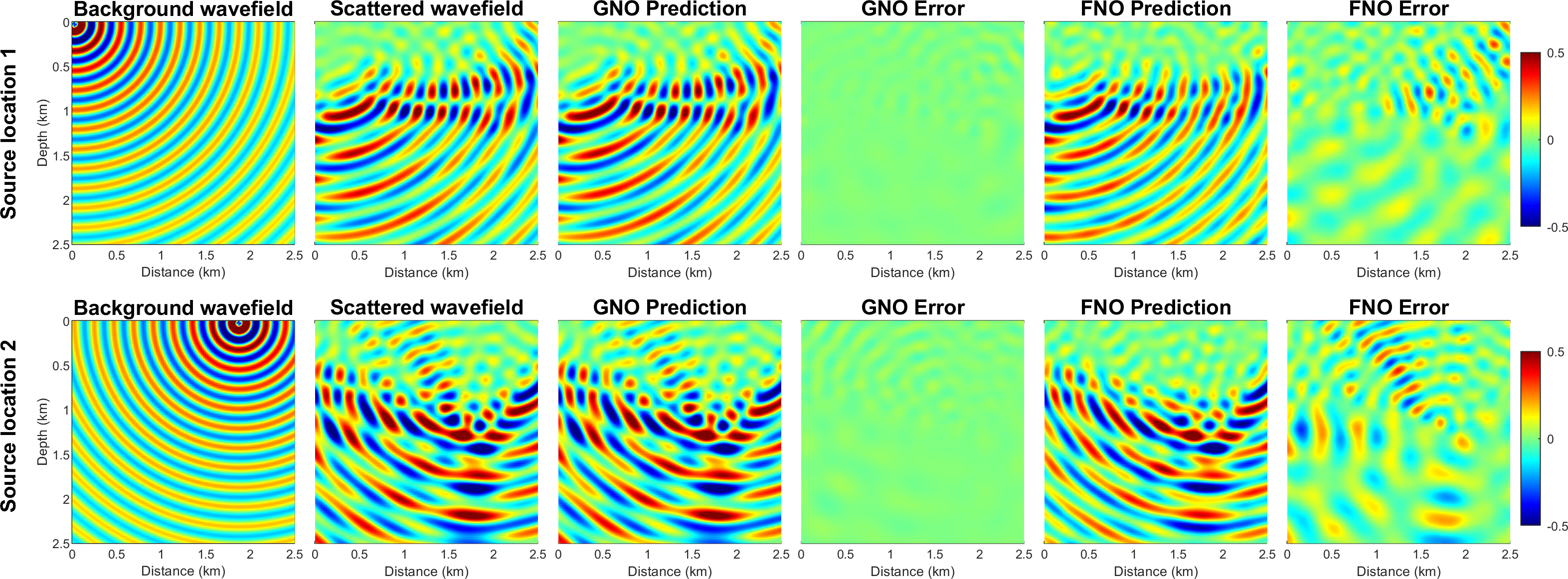}
\caption{Comparison of real-part 8 Hz scattered wavefield predictions for an in-distribution velocity model with two different source locations. The first and second rows correspond to different source positions, and the subfigure ordering within each row follows the same pattern as in Figure~\ref{fig7}.}
\label{fig8}
\end{figure}

To further assess the robustness of GNO at various source positions, we show in Figure \ref{fig8} results for 8 Hz wavefields with two different source locations. The first and second rows correspond to two distinct source positions, while the subfigure ordering within each row follows the same pattern as in Figure~\ref{fig7}. Again, our GNO accurately represents the scattered wavefields for both source positions, maintaining minimal discrepancy with the reference solutions. In contrast, the FNO exhibits more pronounced errors across the domain. Additionally, by inspecting the background wavefields in Figures \ref{fig7} and \ref{fig8}, we can observe how they implicitly encode the frequency and source location information. Consequently, including these background wavefields as part of the input effectively provides the necessary constraints for generating accurate wavefields under varying conditions.

\begin{figure}[htbp]
\centering
\includegraphics[width=1\textwidth]{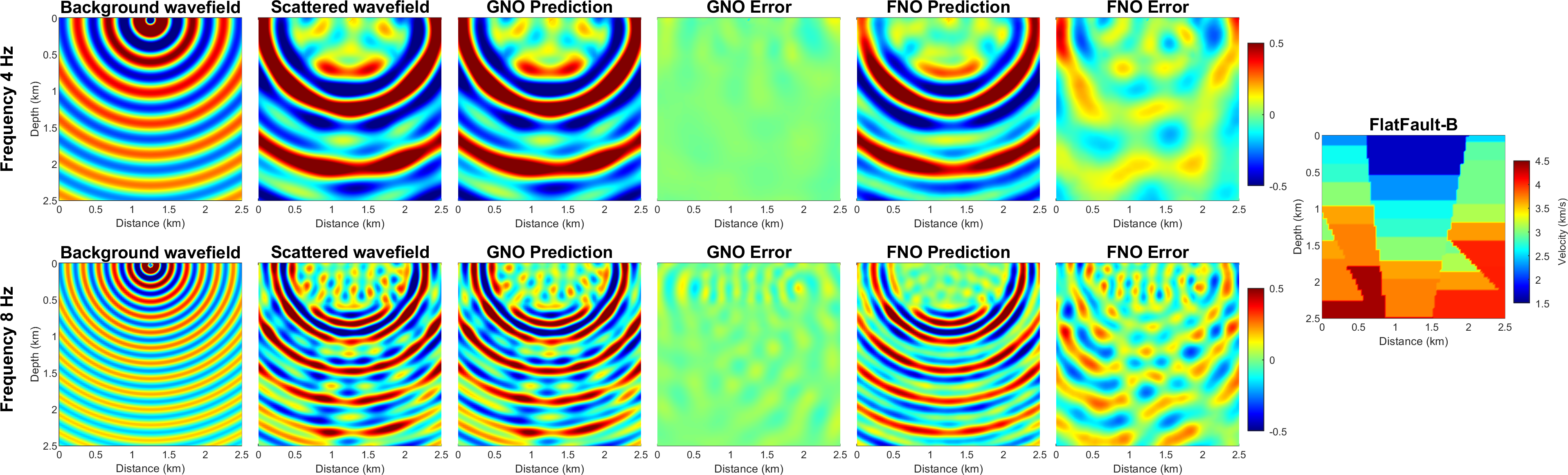}
\caption{Wavefield representations for an out-of-distribution velocity model from FlatFault-B class, at 4 Hz (top row) and 8 Hz (bottom row). For each row we show (from left to right): background wavefield, reference scattered wavefield (numerical solution), GNO prediction, GNO error, FNO prediction, and FNO error. The velocity model is displayed in the far-right panel.}
\label{fig9}
\end{figure}

\begin{figure}[htbp]
\centering
\includegraphics[width=1\textwidth]{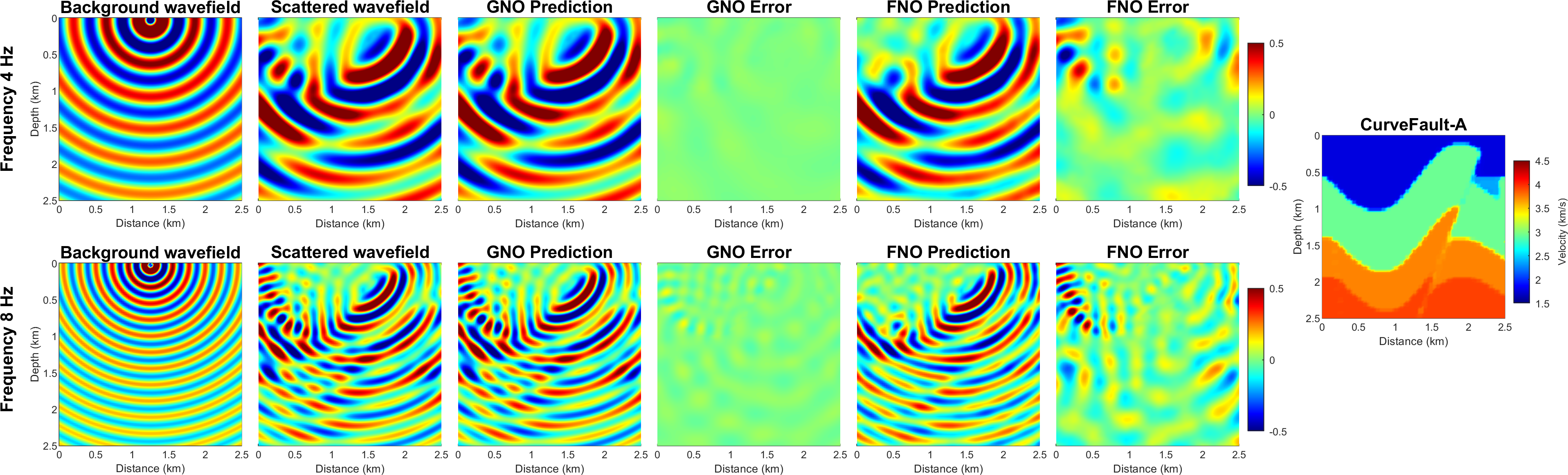}
\caption{Similar to Figure \ref{fig9}, but for the velocity model from CurveFault-A class.}
\label{fig10}
\end{figure}

\begin{figure}[htbp]
\centering
\includegraphics[width=1\textwidth]{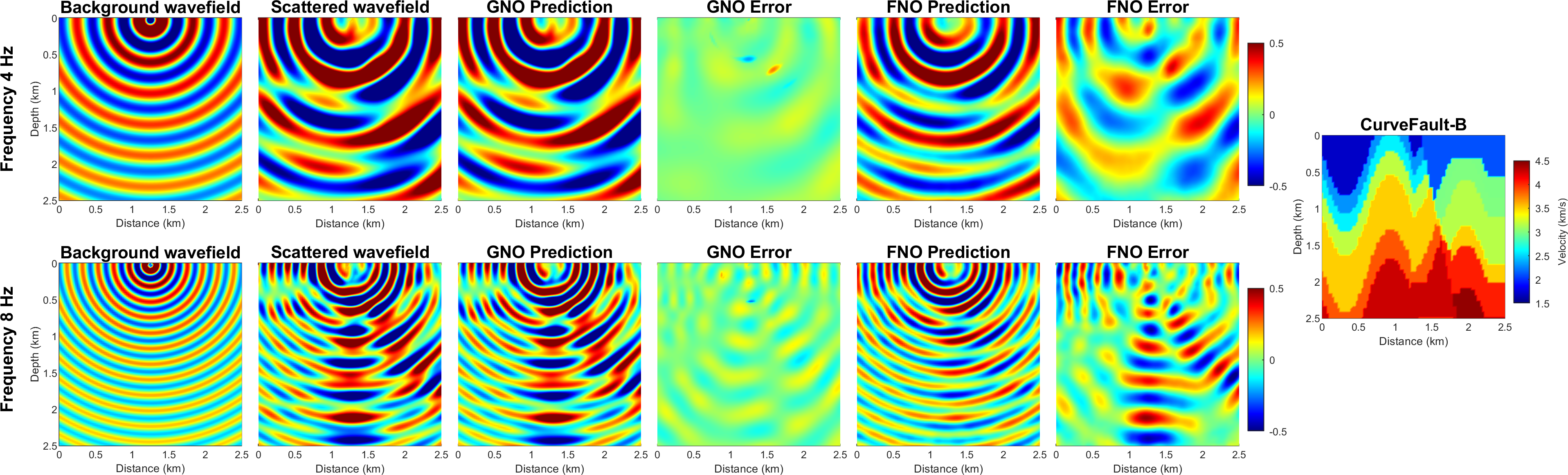}
\caption{Similar to Figure \ref{fig9}, but for the velocity model from CurveFault-B class.}
\label{fig11}
\end{figure}

\subsection{Test on out-of-distribution scenarios}
We further assess the generalization ability of our proposed GNO on velocity models and frequencies outside the training distribution. First, we examine wavefield predictions for three complex velocity models drawn from FlatFault-B, CurveFault-A, and CurveFault-B in the OpenFWI dataset. These model classes are not included in the training set. Figures~\ref{fig9}, \ref{fig10}, and \ref{fig11} show the results for these three models at 4 Hz and 8 Hz, which remain within the frequency range seen during training. In each figure, the top row corresponds to the 4 Hz wavefield and the bottom row to the 8 Hz wavefield. Following the ordering of Figure~\ref{fig7}, each row displays (from left to right) the background wavefield, the reference scattered wavefield, our GNO prediction, the GNO error, the FNO prediction, and the FNO error; the velocity corresponding model is shown on the far right. 

From Figures~\ref{fig9}--\ref{fig11}, we can see that our GNO consistently produces highly accurate scattered wavefields for all three out-of-distribution velocity models at both 4 Hz and 8 Hz. However, the errors are larger for the higher frequency (8 Hz) than for the lower frequency (4 Hz). In contrast, FNO displays significant inaccuracies across both frequencies, particularly for velocity models with more complex structures (e.g., FlatFault-B and CurveFault-B). This discrepancy highlights the limited generalization of FNO in challenging, out-of-distribution scenarios. 

\begin{figure}[htbp]
\centering
\includegraphics[width=1\textwidth]{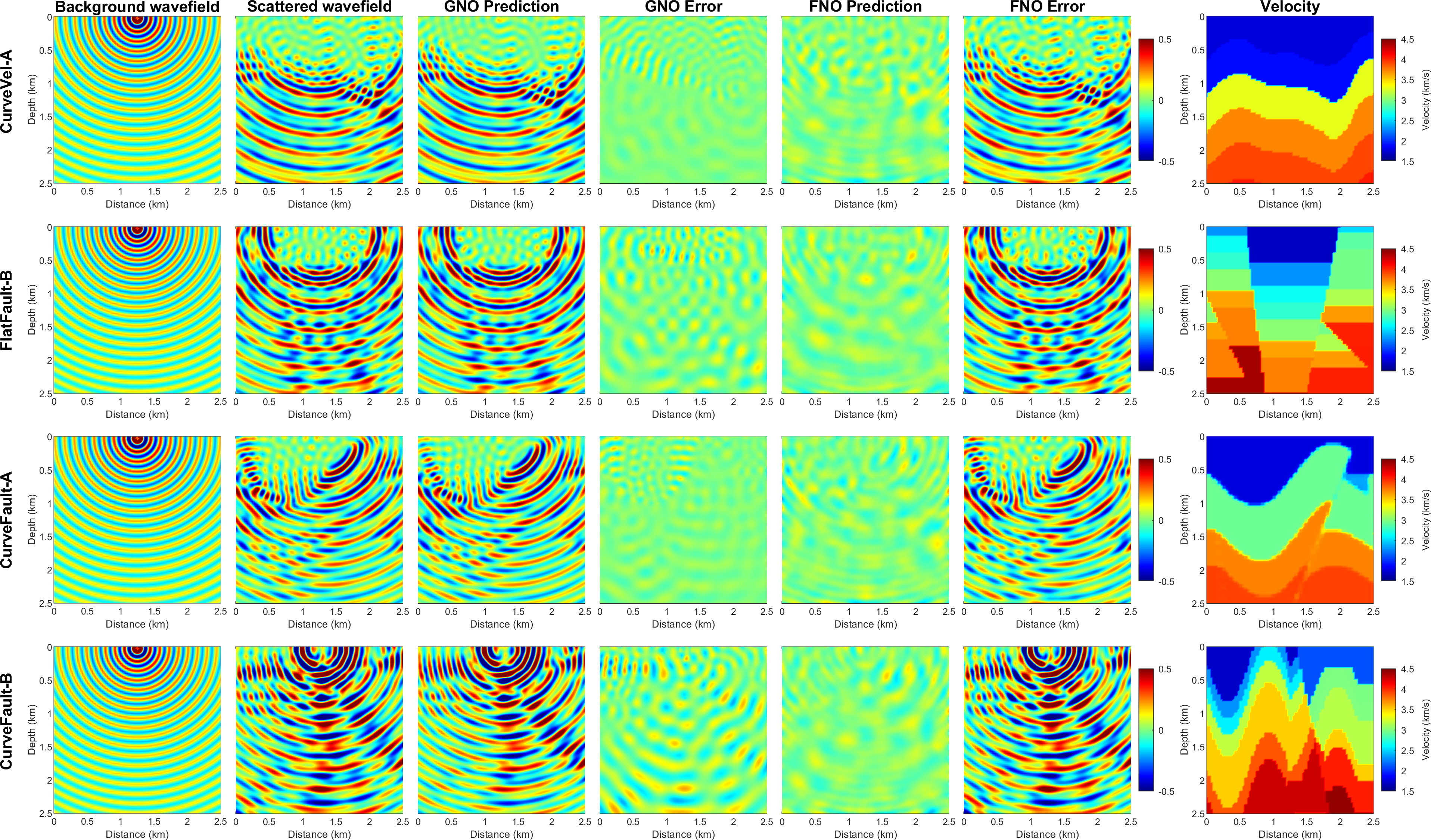}
\caption{Wavefield representations at 12 Hz for four velocity models (from top to bottom): CurveVel-A, FlatFault-B, CurveFault-A, and CurveFault-B. For each row we show from left to right: the background wavefield, reference scattered wavefield, GNO prediction, GNO error, FNO prediction, FNO error, and the corresponding velocity model.}
\label{fig12}
\end{figure}

Next, we test the ability of our GNO to handle frequencies outside the training range by predicting 12 Hz wavefields, as shown in Figure~\ref{fig12}. In addition to an in-distribution velocity model (CurveVel-A), we also include the three out-of-distribution models (FlatFault-B, CurveFault-A, and CurveFault-B) from Figures~\ref{fig9}--\ref{fig11}. Across all four velocity models, our GNO still captures most of the key wavefield features at 12 Hz, but has higher errors than when predicting within the training frequency range. In contrast, FNO fails to produce a reasonable wavefield shape at this unseen frequency, displaying very large errors and underscoring its weaker generalization compared to our GNO. 

\begin{figure}[htbp]
\centering
\includegraphics[width=1\textwidth]{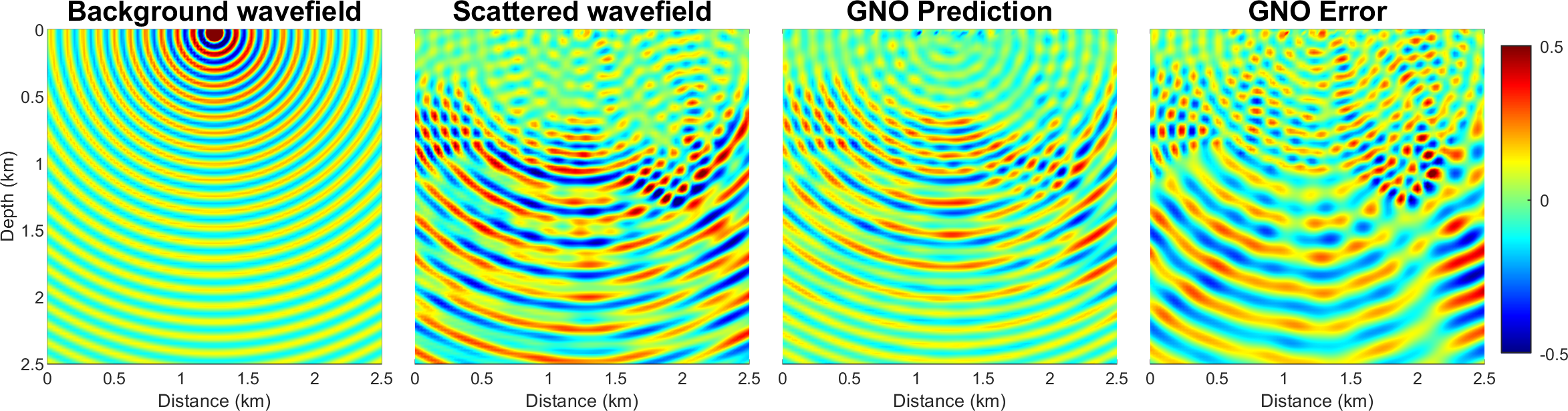}
\caption{Wavefield representations at 15 Hz for the velocity model from CurveVel-A class. From left to right: background wavefield, reference scattered wavefield, GNO prediction, and GNO error.}
\label{fig13}
\end{figure}

However, we emphasis that our GNO’s generalization capability is not without limits. For instance, Figure~\ref{fig13} shows the predicted wavefield at 15 Hz for a CurveVel-A velocity model. While our GNO still produces the broad wavefield structure, the deviation from the numerical reference is substantial. In other words, when the target frequency lies far outside the training distribution, GNO’s accuracy can degrade significantly. Hence, to extend the range of reliable wavefield representations, the training data coverage should be as broad as possible. This observation reinforces the importance of preparing a comprehensive training dataset for enhanced generalization. 

\section{\textbf{Discussion}}
The results of our numerical experiments demonstrate the potential of a diffusion-driven generative neural operator (GNO) for accurately approximating seismic scattered wavefields under a variety of conditions. One of the primary findings is the effectiveness of physics-guided sampling, which injects wave-equation-based corrections during each sampling step. By penalizing the PDE residual at intermediate stages, the model quickly and reliably converges to physically meaningful solutions, mitigating the noise that typically persists for more steps in unguided diffusion processes. This aspect was most evident when observing the rapid decrease in PDE loss and the faster emergence of coherent wavefield structures. 

Another noteworthy advantage of our GNO is its ability to generalize beyond the training distribution. In experiments with velocity models and frequencies not seen in training, the GNO predictions maintained a relatively small error margin, indicating that it learns robust features of the wavefield across different medium complexities and frequency ranges. Although its accuracy does decline at frequencies that lie far outside the training range, its performance still surpasses the baseline Fourier neural operator (FNO), which struggled significantly in these out-of-distribution scenarios. This finding underscores the importance of both the generative diffusion mechanism, which learns a more flexible data manifold, and the physics guidance, which provides consistent constraints even for partially unseen conditions. 

With respect to the cost, like many foundation models, GNO's main cost is involved in the training. In the examples we shared, the training of the diffusion model was slightly higher than a similar training on an FNO. Since the generalization of GNO is superior, we can treat GNO more like a foundation model for scattered wavefields in which the cost can be treated as overhead. On the other hand, we can just use a single sampling step (i.e., direct prediction at the last step) to generate the scattered wavefield and, thus, the inference speed of our GNO to generate wavefields are as efficient as conventional neural operators, like FNO. 

Despite these strengths, an important limitation of our current diffusion-model framework lies in handling large-scale velocity models. As the domain size increases, the model’s ability to capture fine-scale variations in the wavefield become more challenging. However, recent advances in diffusion-model architectures suggest that it is possible to train on lower-resolution data and then generate high-resolution outputs, effectively overcoming such scaling issues \citep{le2024brush}. Additionally, adopting a latent diffusion approach \citep{rombach2022high} could compress high-dimensional wavefield data into a lower-dimensional latent space, reducing memory usage and computational cost while preserving essential wavefield features. By integrating these newer methods, our GNO could achieve more accurate large-scale wavefield solutions without substantially increasing computational overhead.
\section{\textbf{Conclusions}}
We proposed a generative neural operator (GNO) that leverages generative diffusion models (GDMs) to represent complex scattered wavefields. Unlike traditional approaches that rely on direct numerical solvers, our method was trained on scattered wavefields corresponding to a broad range of velocity models, frequencies, and diverse source locations. Through this training, GDMs capture the underlying statistical distribution of the scattered wavefields, thereby allowing us to treat wavefield generation as a learning-based, generative modeling task. As a result, new wavefields can be rapidly synthesized at the inference stage, constructing a highly efficient and generalizable neural operator. To further ensure physical consistency, we introduced a physics-guided sampling procedure that penalizes PDE residuals at each generation step. By continuously forcing the intermediate wavefields toward solutions of the governing Helmholtz equation, this procedure produces more accurate and stable results than purely data-driven diffusion. Numerical examples demonstrated that our GNO, equipped with physics guidance, yields highly accurate wavefield representations for in-distribution scenarios while also exhibiting enhanced generalization to unseen velocity models and frequencies. This improvement is particularly evident when the model is applied to out-of-distribution cases, where conventional neural operators often fail to capture crucial wavefield features. Thus, by combining the flexibility of GDMs with explicit physical constraints, our GNO achieves efficient, robust, and scalable seismic wavefield representation.
\section{\textbf{Acknowledgment}}
This publication is based on work supported by the King Abdullah University of Science and Technology (KAUST). The authors thank the DeepWave sponsors fort heir support. This work utilized the resources of the Supercomputing Laboratory at King Abdullah University of Science and Technology (KAUST) in Thuwal, Saudi Arabia.
\vspace{0.5cm}
\section{\textbf{Code Availability}}
The data and accompanying codes that support the findings of this study are available at: 
\\
\url{https://github.com/DeepWave-KAUST/GNO}. (During the review process, the repository is private. Once the manuscript is accepted, we will make it public.)

\bibliographystyle{unsrtnat}
\bibliography{references}

\end{document}